\documentclass[a4paper,11pt]{article}
\usepackage{jcappub} % for details on the use of the package, please see the JINST-author-manual
\usepackage{lineno}
\usepackage{CJK}
\usepackage{xcolor}

\newcommand{\md}{\mathrm{d}}
\newcommand{\me}{\mathrm{e}}

\title{\boldmath Constraints on Axion-like Dark Matter from Cosmic Birefringence with a Polarization Array of Repeating Fast Radio Bursts}

\author[a,b]{Xiaohui Liu}
\author[c]{Zi-Yan Yuwen}
\author[d]{Yun-Long Zhang}
\author[d,b]{Zerui Liu}
\author[e]{Shiqian Zhao}
\author[d,b]{Shuai Feng}
\author[b]{Wei-Yang Wang}
\author[a,b]{Xuelei Chen}
\affiliation[a]{State Key Laboratory of Radio Astronomy and Technology, National Astronomical Observatories, CAS, A20 Datun Road, Chaoyang District, Beijing, 100101, P. R. China}
\affiliation[b]{School of Astronomy and Space Science, University of Chinese Academy of Sciences, Beijing 100049, People's Republic of China}
\affiliation[c]{Asia Pacific Center for Theoretical Physics (APCTP), Pohang 37673, Korea}
\affiliation[d]{National Astronomical Observatories, Chinese Academy of Sciences, 20A Datun Road, Chaoyang District, Beijing 100101, People's Republic of China}
\affiliation[e]{Department of Astronomy, School of Physics and Materials Science, Guangzhou University, Guangzhou, 510006, China}
\emailAdd{wywang@ucas.ac.cn, xuelei@cosmology.bao.ac.cn}

\abstract{Axion-like dark matter (ALDM) is a well-motivated dark matter candidate whose Chern-Simons coupling to photons can induce cosmic birefringence, producing temporal oscillations and spatial correlation in polarization angle (PA) signals from different sources. Hyperactive repeating fast radio bursts (FRBs) with high linear polarization are a particularly promising probe of such signatures. We develop a Bayesian framework to search for PA oscillations caused by ALDM in individual repeating FRBs, and combine multiple sources in an FRB polarization array (FRBPA) by exploiting correlations arising from the common Earth term. We apply this framework to PA datasets from three well-studied hyperactive repeating FRBs, and constrain the Chern-Simons coupling over a broad range of ALDM masses. For each FRB, we estimate the dark matter density at its host location using a simulation-calibrated halo profile. The inferred dark matter densities are comparable to the fiducial local density $\rho_0 = 0.4~\mathrm{GeV/cm^3}$. The comparable source and Earth dark matter densities, together with the redshift-induced difference between the frequencies of the source and Earth terms, motivate our adoption of the full two-component birefringence model. We find no compelling evidence for ALDM-induced PA oscillations in either the individual source analyses or the joint FRBPA analysis. We place 95\% credible upper limits on the Chern-Simons coupling strength as a function of ALDM mass. Across the axion mass from $10^{-23}$ to $10^{-18}~\mathrm{eV}$, the resulting upper limits of the coupling strength of FRBPA lie between $10^{-12.6}$ and $10^{-7.9}~\mathrm{GeV}^{-1}$. Our results establish long-term polarimetric monitoring of hyperactive repeating FRBs as a promising and independent probe of ALDM, with sensitivity expected to improve with longer observational baselines and larger source samples.}

\begin{document}
\maketitle
\flushbottom

\section{Introduction}
\label{sec:intro}
General relativity has achieved great success in explaining the universe from the Solar System to cosmological scales~\citep{1973Sci...182..229B, 1979Natur.277..437T, 2022ApJ...927...28L, 2025A&A...704A.153M, 2026arXiv260719293T}. Within the standard cosmology, the energy budget is dominated by the mysterious dark components: dark energy and dark matter (DM) \citep{2020A&A...641A...6P}. Dark energy is believed to drive the accelerated expansion of the universe, while DM dominates the structure formation and evolution. Despite its pivotal role in shaping the cosmos, the nature of DM remains one of the most perplexing puzzles in modern physics.

The concept of axion-like dark matter (ALDM) arises naturally in many extensions of the Standard Model and has become one of the most widely studied DM candidates \citep{1990PhRvD..41.1231C, 2000PhRvL..85.1158H, 2016PhR...643....1M, 2017PhRvD..95d3541H, 2025PhyR.1117....1C}.
Existing constraints on ALDM can be broadly classified into gravitational probes and non-gravitational probes.
Gravitational probes mainly rely on the characteristic suppression of small-scale structures caused by the quantum pressure of the ultralight ALDM. For instance, the abundance of Milky Way satellite galaxies can place a lower bound on the ALDM mass at $2.9 \times 10^{-21}$ eV \citep{2021PhRvL.126i1101N}. The most stringent constraints from this class of probes come from the Lyman-$\alpha$ forest \citep{2021PhRvL.126g1302R} and ultrafaint dwarfs \citep{2021ApJ...912L...3H, 2022PhRvD.106f3517D}, which set lower limits at approximately $10^{-20} - 10^{-19}$ eV and appear to disfavor the parameter space of the canonical fuzzy DM mass region. However, other studies have pointed out that systematic uncertainties and mixed DM models might weaken this constraint \citep{2018ApJ...863...73Z, 2021A&ARv..29....7F, 2024MNRAS.530.4920K, Liu:2026jkq}.

On the other hand, one can also constrain the axion-photon coupling constant $g_{a\gamma\gamma}$ at different axion masses to limit the properties of ALDM.
Axion–photon conversion in magnetic fields can produce additional photon signals or distort observed photon spectra, providing constraints on $g_{a\gamma\gamma}$ over a broad range of axion masses \citep{2017NatPh..13..584C, 2015JCAP...02..006P, 2022MNRAS.510.1264S, 2025ApJ...988..104Y, 2025PhRvD.111l3010T, 2026A&A...707A.108M, 2026JCAP...03..053F}.
If ALDM couples to photons through the Chern-Simons interaction, it can affect the propagation of linearly polarized electromagnetic waves.
This leads to an additional phase shift between the left and right circular polarization modes when the linearly polarized light propagates through the ALDM field. This effect, known as the cosmic birefringence effect, predicts an oscillating polarization angle (PA) background in addition to the intrinsic PA of astronomical light, which provides a novel way to detect the presence of ALDM.
This effect can be searched for through long-term monitoring of linearly polarized astrophysical sources, and a series of interesting results has been obtained \citep{2019PhRvD.100a5040F, 2024PhRvD.110f3013A, 2024PhRvD.110j3525Z, 2025PhRvD.112l3513N, 2026PhRvD.113k5036J, 2026PhRvD.113d3044A, 2026PDU....5202260A, Wang:2026bns, Qin:2026hkx, 2026arXiv260215611N}.
Recently, pulsars have been proposed as probes of the oscillating PA background induced by ALDM, as they typically exhibit significant linear polarization and a predictable time-averaged PA profile \citep{2023PhRvL.130l1401L}. Thus, the deviations from the expected PA profile can be used to give constraints on $g_{a\gamma\gamma}$ as a function of axion mass \citep{2022JCAP...06..014C, 2025PhRvD.111f2005P}. Since most pulsars reside within the Milky Way, their separations are generally comparable to the coherence length of the fuzzy DM field. As a result, the spatial correlation between them can be used to form a Pulsar Polarization Array (PPA)~\citep{Liu:2021zlt} to further improve the constraining power and reach the tightest constraints till now \citep{2026PhRvL.136a1001X, 2026PhRvD.113d3059L, 2026arXiv260529106S, Yuwen:2026zjk}.
Close binaries have also been proposed as targets for a similar method \citep{2026arXiv260704550M}.

Nevertheless, the low flux densities of pulsars, the need for pulse averaging, and additional noise and systematics motivate searches using brighter linearly polarized sources. Fast radio bursts (FRBs) are exceptionally bright radio transients detectable at cosmological distances \citep{2007Sci...318..777L, 2013Sci...341...53T, 2023RvMP...95c5005Z}. Bursts from some repeating sources exhibit nearly 100\% linear polarization \citep{2018Natur.553..182M, 2020ApJ...896L..41C}, providing sufficiently bright polarized emission for precise measurements of both the rotation measure (RM) and PA from individual bursts.
Indeed, gravitationally lensed FRBs have already been proposed as probes of Galactic ALDM through differential birefringence measurements \citep{2024PhRvD.109b1303G}.
Applying birefringence searches to repeating FRBs requires accounting for two important features. First, their PAs can exhibit substantial scatter between bursts \citep{2021MNRAS.508.5354H, 2026arXiv260500372L, 2026arXiv260428012M, 2026arXiv260216409U}, even when they remain nearly constant within individual burst profiles \citep{2018ApJ...863....2G, 2019ApJ...885L..24C, 2021NatAs...5..594N, 2021ApJ...908L..10H, 2022ApJ...932...98S, 2023MNRAS.524.3303B, 2023ApJ...950...12M, 2023MNRAS.526.3652K, 2024ApJ...968...50P, 2025ApJ...988..175L}. This additional scatter must be accounted for when searching for ALDM signals in PA time series. Second, FRBs in distinct host galaxies are separated by distances far exceeding the ALDM coherence length over the fuzzy mass range considered here, allowing their source terms to be treated as statistically independent. In this regime, correlations between the ALDM-induced PA variations of different sources arise solely from the common Earth term, enabling the construction of an FRB Polarization Array (FRBPA).

In this work, we develop a Bayesian framework that explicitly models intrinsic PA scatter and incorporates both the independent source terms and the common Earth term. We apply this framework to the Five-hundred-meter Aperture Spherical radio Telescope (FAST) \citep{2011IJMPD..20..989N, 2018IMMag..19..112L, 2019SCPMA..6259502J} observations of three hyperactive repeating FRBs, analyzing each source individually and combining them into an FRBPA to constrain the ALDM–photon coupling.
We validate the method and present the constraints from an FRBPA for the first time.
Longer observational baselines are particularly valuable for resolving the slow PA oscillations associated with low-mass ALDM. The current datasets have a longest individual-source baseline of approximately 472 days. Consequently, the results presented here should be regarded as a proof-of-concept demonstration of the method rather than the ultimate constraint. As monitoring campaigns continue and larger PA samples accumulate over longer time baselines, substantially improved sensitivity to low-mass ALDM is expected.

This paper is organized as follows. In Sec. \ref{sec:bayes}, we describe the theoretical basics of our hierarchical Bayesian inference. In Sec. \ref{sec:dm}, we describe how we estimate the DM density at each FRB location using a simulation-calibrated halo model. In Sec. \ref{sec:res}, we present our main results and compare them with other observations. The conclusions are given in Sec. \ref{sec:con}. 

\section{Inference Framework} \label{sec:bayes}
In this section, we present our hierarchical Bayesian framework for using the PA measurements of repeating FRBs to probe ALDM.

Each PA curve in our sample is described by the peak arrival time $t$ in Barycentric Coordinate Time (TCB), the mean PA $\psi$, and its corresponding uncertainty $\sigma_{\psi}$, collectively denoted by $\boldsymbol{\theta} = \{ t, \psi, \sigma_{\psi} \}$.
The reference frequency of the arrival time has been uniformly converted to 1.5 GHz for all bursts, allowing different samples to be combined consistently in the FRBPA method.

\subsection{ALDM induced oscillating PA background} \label{subsec: ALDM}
As the cosmic birefringence effect is topological \citep{1992PhLB..289...67H}, the ALDM-induced PA shift only depends on the initial and final DM density, which can be written as
\begin{equation} \label{eq:a-a}
    \Delta \psi^a = \frac{g_{a\gamma\gamma}}{2} \left[ a\left(\vec{x}_s, t_s\right) - a\left(\vec{x}_0, t_0\right) \right],
\end{equation}
where $a\left(\vec{x}_s, t_s\right)$ and $a\left(\vec{x}_0, t_0\right)$ represent the ALDM field at the source and Earth positions, and $g_{a\gamma\gamma}$ is the crucial parameter that characterizes the Chern-Simons coupling strength and is the main target we aim to constrain. 

On galactic scales, the axion field can be approximated as a stochastic superposition of harmonic oscillations \citep{2023PhRvD.108i2010N, 2021PhRvD.103g6018F},
\begin{align}
    a(\vec{x},t) = \frac{\sqrt{2\rho(\vec{x})/N_a}}{m_a} \sum_{i=1}^{N_a} \cos\left( \omega_i t - \vec{k}_i\cdot \vec{x} + \phi_i \right)~,
\end{align}
where $\rho(\vec{x})$ is the local ALDM density, and the index $i$ runs from $1$ to the particle occupation number $N_a$. For the $i$-th particle, the random phase $\phi_i$ follows a uniform distribution $U[0,2\pi]$.
For convenience, we adopt natural units with the speed of light and the reduced Planck constant set to unity in the following derivation.
In the non-relativistic limit $v_i\ll 1$, the angular frequency reduces to the mass of the particle $\omega_i\simeq m_a$. At a fixed position, the spatial phase originating from the momentum $k_i$ can be effectively absorbed in a random phase by redefining $\phi_i \to \phi_i - \vec{k}_i\cdot \vec{x}$. In the large-occupation-number limit $N_a\to \infty$,
\begin{align}
    a(\vec{x},t) = \frac{\sqrt{2\rho(\vec{x})}}{m_a} \mathrm{Re}\left( \me^{im_a t} \sum \sqrt{f(\vec{v})\Delta v^3} \, \me^{i\phi_{\vec{v}}} \right)~ \equiv \frac{\sqrt{2\rho(\vec{x})}}{m_a} \mathrm{Re}\left( \me^{im_a t} Z\right),
\end{align}
where $f(\vec{v})$ denotes the velocity distribution of the ALDM satisfying normalization condition $\sum f(\vec{v})\Delta v^3=1$, while $Z$ is a complex random variable encoding the stochastic superposition of the velocity modes. By the central limit theorem, $Z$ follows a circular complex Gaussian distribution with $\langle Z\rangle = 0$ and $\langle |Z|^2\rangle = 1$. Therefore, $Z$ can be parameterized as $Z=\alpha \exp(i\phi)$~\citep{2018PhRvD..97l3006F, 2021NatCo..12.7321C} with an amplitude $\alpha$ following the Rayleigh distribution $P(\alpha) = 2\alpha \exp{( -\alpha^2 )}$ and a phase following uniform distribution $U[0,2\pi]$, yielding
\begin{align}\label{eq: axion field at given location}
    a(\vec{x},t) = \frac{\sqrt{2\rho(\vec{x})}}{m_a} \, \alpha \cos(m_a t + \phi).
\end{align}

It should be noted that, for a source located at a cosmological redshift $z$, two neighboring emission and observation events of photons are related by the cosmological time dilation $\md t_s=\md  t_0/(1+z)$. Consequently, a source-frame axion oscillation $\cos(m_a t_s+\phi)$ appears to the observer as $\cos\left(m_a t_0/(1+z)+\phi\right)$, where the time offset associated with the cosmological propagation has been again absorbed into the random phase. Thus, from the observer's perspective, the source term oscillates at the redshifted angular frequency $\omega_s = m_a/(1+z)$. Incorporating this redshift and substituting Eq.~\eqref{eq: axion field at given location} back into Eq.~\eqref{eq:a-a} results in
\begin{align}\label{eq:aldm}
    \Delta \psi^a(t) = \frac{g_{a\gamma\gamma}}{\sqrt{2} m_a} \left( \alpha_s \sqrt{\rho_s} \cos \left(\frac{m_a t}{1+z} + \phi_s\right) - \alpha_0 \sqrt{\rho_0} \cos (m_a t + \phi_0) \right).
\end{align}
We adopted the local DM density $\rho_0 = 0.4 ~\mathrm{GeV / cm^3}$ determined from the measurement of the Milky Way rotation curve \citep{2010A&A...523A..83S}. As for the term $\rho_s$, we estimate it using the simulation-calibrated generalized Navarro-Frenk-White (gNFW) DM profile, and the details are presented in Sec. \ref{sec:dm}.

\subsection{Likelihood}
Due to the substantial intrinsic burst-to-burst scatter of the PA of repeating FRBs, we must simultaneously infer the properties of axions as well as the population-level properties. A hierarchical Bayesian framework provides a natural approach for simultaneously modeling these contributions.
To determine the mean PA of each burst, we employ a Markov Chain Monte Carlo method and obtain 2000 posterior samples of $\psi$ from $p(\boldsymbol{\theta} | x)$. We then approximate the posterior distribution by a Gaussian distribution characterized by $\psi$ and $\sigma_{\psi}$.
Since a flat prior on $\psi$ and $\sigma_{\psi}$ is adopted in the estimation, the likelihood is proportional to the posterior, which means that $\mathcal{L}(x | \boldsymbol{\theta}) \propto p(\boldsymbol{\theta} | x)$. The resulting parameters $\boldsymbol{\theta}$ are used as the input to subsequent population-level analysis.

The population distribution $p_{\mathrm{pop}}(\boldsymbol{\theta} | \boldsymbol{\Lambda})$ is introduced to describe the intrinsic PA distribution and is parameterized by a set of hyperparameters $\boldsymbol{\Lambda} = \{ \boldsymbol{\Lambda}^p, \boldsymbol{\Lambda}^a \}$, where $\boldsymbol{\Lambda}^p$ and $\boldsymbol{\Lambda}^a$ denote the population and ALDM-related parameters, respectively.
The posterior distribution on hyperparameters given the $N$ PA curves in the sample can be written as
\begin{equation}
    p(\boldsymbol{\Lambda} | \{x\}) \propto p(\boldsymbol{\Lambda}) \prod_{i=1}^{N} \int \mathcal{L}(x_i | \boldsymbol{\theta}) p_{\mathrm{pop}}(\boldsymbol{\theta}| \boldsymbol{\Lambda}) d \boldsymbol{\theta},
\end{equation}
where $\{x\}$ denotes the set of observed PA curves, $p(\boldsymbol{\Lambda})$ denotes the prior of the hyperparameters $\boldsymbol{\Lambda}$, and $\mathcal{L}(x_i | \boldsymbol{\theta}, \boldsymbol{\Lambda})$ is the likelihood of a single PA curve. Following the previous work \cite{2026arXiv260500372L}, we use a time-independent Gaussian model to describe the population behavior, which is expressed by
\begin{equation}
    p_{\mathrm{pop}} (\boldsymbol{\theta} | \boldsymbol{\Lambda}) = \frac{1}{\sqrt{2 \pi \sigma^2}} \exp \left( -\frac{(\psi-\mu)^2}{2\sigma^2} \right),
\end{equation}
where $\mu$ is the location where the PA is clustered around and $\sigma$ is the width. For our null model, the location $\mu$ is time-independent, so it can be treated as a constant $\mu = \mu_A$, where A denotes a specific repeating FRB source. As for the scenario where ALDM exists, the oscillating PA background from the cosmic birefringence effect would give rise to an extra time-dependent component,  $\mu = \mu_A + \Delta \psi^a(t)$, as described in Sec. \ref{subsec: ALDM}.

It should be noted that the PA is periodic over $\pi$. To avoid biases associated with the angular boundary, the PA measurements of each source are centered with respect to the corresponding population mean, and the resulting residuals are mapped onto the principal interval $[-\pi/2,\pi/2)$. The likelihoods below are therefore evaluated in terms of zero-centered PA residuals. For the individual-source analysis, the population mean includes both $\mu_A$ and the time-dependent ALDM contribution, whereas in the FRBPA analysis, the ALDM contribution is encoded in the covariance matrix.

\subsubsection{Individual repeating FRBs}
For an individual repeating FRB, the ALDM-induced PA background is treated as a deterministic realization of the source and Earth ALDM field. The model PA is given by $\mu = \mu_A + \Delta \psi^a(t)$.
Under the Gaussian approximation for the burst-level PA posterior, the convolution between the measurement uncertainty and the population distribution can be performed analytically. 
Consequently, the effective covariance of the noise is $C^n_{ij} = \delta_{ij} (\sigma^2_{\psi,i}+\sigma^2)$. The final log-probability can be written as
\begin{equation}\label{eq:l1}
    \ln p(\boldsymbol{\Lambda} | \{x\}) \propto \ln p(\boldsymbol{\Lambda}) - \frac{1}{2} \boldsymbol{\Delta \psi}^{T} \boldsymbol{C}^{-1} \boldsymbol{\Delta \psi} - \frac{1}{2} \ln |2 \pi \boldsymbol{C}|,
\end{equation}
where $p(\boldsymbol{\Lambda})$ is the prior, and $\boldsymbol{\Delta \psi} = \boldsymbol{\psi} - \mu(t)$ is the input PA vector.

\subsubsection{FRB polarization array}
We further combine the PA measurements from multiple repeating FRBs to construct an FRBPA. For $N$ sources, the complete PA vector is formed by concatenating the $N$ source-specific PA vectors, $\boldsymbol{\psi} = (\boldsymbol{\psi}_1, ..., \boldsymbol{\psi}_n)^T$.
Unlike the individual-source analysis, the random amplitudes and phases of the ALDM fields are marginalized analytically in the FRBPA analysis. Consequently, the ALDM signal is represented by an additional covariance matrix rather than by a deterministic contribution to $\mu$.
The covariance matrix of the axion contribution $C^a_{A,n;B,m}$ can be calculated by
\begin{equation}\label{eq: def Ca}
\begin{aligned}
    C^a_{A,n;B,m} &= \langle \Delta \psi^a_{A,n} \Delta \psi^a_{B,m} \rangle  \\
    &= \frac{g_{a\gamma\gamma}^2}{2m_a^2} 
    \left\langle \left( \alpha_A \sqrt{\rho_A} \cos \left(\frac{m_a t_n}{1+z_A} + \phi_A \right) - \alpha_0 \sqrt{\rho_0} \cos (m_a t_n + \phi_0) \right) \right .\\
    &\quad\quad\quad\quad \left. \left( \alpha_B \sqrt{\rho_B} \cos \left(\frac{m_a t_m}{1+z_B} + \phi_B\right) - \alpha_0 \sqrt{\rho_0} \cos (m_a t_m + \phi_0)\right) \right\rangle \\
    & = \frac{g_{a\gamma\gamma}^2}{4m_a^2} \left( \rho_0 \cos\left(m_a(t_n-t_m)\right) + \delta_{AB}\,  \rho_A\cos\left(\frac{m_a (t_n-t_m)}{1+z_A}\right) \right),
\end{aligned}
\end{equation}
where $A$ and $B$ labels different FRB sources, and $t_n \,(t_m)$ is the $n$-th ($m$-th) burst for the $A$-th ($B$-th) repeating FRB.
In the last line, we have used the fact that $\langle \alpha_A \alpha_B \cos(x+\phi_A) \cos(y+\phi_B) \rangle = \frac{1}{2}\cos(x-y)\delta_{AB}$.
For $A\neq B$, the term vanishes upon averaging over the independent and uniformly distributed phases. For $A=B$, averaging over the common phase gives $\frac{1}{2}\cos(x-y)$, and we have used the normalization $\langle\alpha^2\rangle=1$.

One major difference between Eq.~\eqref{eq:aldm} and the corresponding ALDM-induced PA shift in a pulsar polarization array is that the FRB sources and the Earth reside in distinct DM halos. In a pulsar polarization array, both the pulsars and the Earth reside within the Milky Way halo, and their ALDM fields may retain spatial correlations on scales comparable to the ALDM coherence length. By contrast, in an FRBPA, the ALDM fields in the host galaxy halos and the Milky Way halo can be treated as statistically independent, allowing the field distributions at the two endpoints to be averaged independently, resulting in no spatial-correlations in the covariance matrix. As a result, correlations between different FRBs arise only through the temporal correlation of the common Earth term, corresponding to the first term in the last line of Eq.~\eqref{eq: def Ca}.

The noise covariance matrix of the FRBPA is given by
\begin{equation}
    C^n_{A,n;B,m} = \delta_{AB}\delta_{nm} \left( \sigma_{\psi,A,n}^{2} + \sigma_A^{2} \right),
\end{equation}
where $\sigma_{\psi,A,n}$ is the measurement uncertainty and $\sigma_A$ represents the intrinsic PA scatter of source $A$. We define the residual PA vector as
\begin{equation}
    \boldsymbol{\Delta\psi} = \boldsymbol{\psi} - \boldsymbol{\mu}.
\end{equation}
For each source A, the corresponding mean $\mu_{A}$ is repeated for all bursts from that source, and the resulting source-specific vectors are concatenated to form $\boldsymbol{\mu}$. The total covariance matrix is
\begin{equation}
    \boldsymbol{C} = \boldsymbol{C}^{n} + \boldsymbol{C}^{a}.
\end{equation}
The log-posterior for an FRBPA containing $N$ sources is then given by
\begin{equation} \label{eq:l2}
    \ln p(\boldsymbol{\Lambda}|\{x\}) \propto \ln p(\boldsymbol{\Lambda}) - \frac{1}{2} \boldsymbol{\Delta\psi}^{T} \boldsymbol{C}^{-1} \boldsymbol{\Delta\psi} - \frac{1}{2} \ln\left|2\pi\boldsymbol{C}\right|.
\end{equation}
We consider two treatments of the inter-source correlations. In the full-correlation analysis, the complete covariance matrix derived above is used, including the off-diagonal covariance blocks generated by the common Earth term. In the auto-correlation analysis, only the covariance between bursts from the same FRB source is retained, and the covariance elements with $A\neq B$ are set to zero. Comparing the two treatments allows us to assess the contribution of the common Earth term to the ALDM constraints.

\subsection{Priors and Bayes factor}
The parameters considered in this work fall into two categories: population parameters and ALDM-related parameters. For each repeating FRB, the population parameters include a time-independent mean PA $\mu_A$ and an intrinsic PA scatter $\sigma$. Both quantities are treated as free parameters and assigned broad uniform priors.
In the individual-source analysis, the ALDM-induced PA signal is treated as a deterministic field realization. The random amplitudes and phases follow Rayleigh distribution and uniform distribution, respectively. In the FRBPA analysis, the random ALDM amplitudes and phases are analytically marginalized and therefore do not appear as explicit sampling parameters. Their statistical effects are instead incorporated into the ALDM covariance matrix $\boldsymbol{C}^{a}$. Because $g_{a\gamma\gamma}$ can span several orders of magnitude, we adopt a log-uniform prior for this parameter.
The prior distribution of the source DM density $\rho_A$ is assumed to follow a lognormal function, and the corresponding parameters of the source A can be determined from the simulations described in Sec.~\ref{sec:dm}. The adopted prior distributions are summarized in Table~\ref{tab:prior1}.

\begin{table}[htbp]
\centering
\begin{tabular}{l|c}
\hline\hline
Parameter & Prior \\
\hline
$\mu_1$, $\mu_2$, $\mu_3$ & Uniform in $[-\pi/2,\pi/2]$ \\
$\sigma_1$, $\sigma_2$, $\sigma_3$ & Uniform in $[0,\pi/2]$ \\
$g_{a\gamma\gamma} \ (\mathrm{GeV}^{-1})$ & LogUniform in $[10^{-15},10^{-5}]$ \\
$\alpha_0$, $\alpha_s$ & Rayleigh distribution \\
$\phi_0$, $\phi_s$ & Uniform in $[0,2\pi]$ \\
$\rho_1~(\mathrm{GeV/cm^3})$ & Lognormal $\mathcal{LN}$(0.78, 0.76) \\
$\rho_2~(\mathrm{GeV/cm^3})$ & Lognormal $\mathcal{LN}$(-0.35, 0.50) \\
$\rho_3~(\mathrm{GeV/cm^3})$ & Lognormal $\mathcal{LN}$(-0.26, 0.52) \\
\hline\hline
\end{tabular}
\caption{Prior distributions adopted for the population and ALDM-related
parameters. The amplitude and phase parameters are explicitly sampled only in the individual-source analysis, whereas they are analytically marginalized in the FRBPA analysis. The subscripts 1, 2, and 3 correspond to FRB 20201124A, FRB 20220912A, and FRB 20240114A, respectively.}
\label{tab:prior1}
\end{table}

In Bayes' theorem, the preference of the data for one model over another is quantified by the Bayes factor (BF), which is defined as the ratio of their corresponding evidences
\begin{equation}
    \mathrm{BF} = \frac{\mathcal{Z}_1}{\mathcal{Z}_0},
\end{equation}
where $\mathcal{Z}_1$ and $\mathcal{Z}_0$ represent the Bayesian evidences of Models 1 and 0, respectively. In this paper, the BF is introduced to evaluate the preference of the ALDM model over the null model. The value of BF for a given model comparison can be interpreted as evidence against or in favor of the ALDM model according to the Jeffreys scale \citep{jeffreys1998theory}. BF$<$1 means that the ALDM scenario is disfavored, while BF values in the ranges [$10^{0.0}$,$10^{0.5}$], [$10^{0.5}$,$10^{1.0}$], [$10^{1.0}$,$10^{1.5}$], [$10^{1.5}$,$10^{2.0}$], and [$10^{2.0}$,$\infty$) are interpreted as negligible, substantial, strong, and decisive evidence of the ALDM model, respectively.
For a given mass $m_a$ of ALDM, we perform a Bayesian analysis to constrain $g_{a\gamma\gamma}$ with the marginalized likelihoods in Eqs. \eqref{eq:l1} and \eqref{eq:l2}, and calculate the BF using the \texttt{bilby} \citep{2019ApJS..241...27A} python package with the \texttt{dynesty} \citep{2020MNRAS.493.3132S} sampler.

\subsection{FRB datasets}
In this paper, we use four large FRB PA datasets from three well-known repeating FRB sources, FRB 20201124A \citep{2022Natur.609..685X, 2022RAA....22l4001Z, 2022RAA....22l4003J, 2022RAA....22l4002Z, 2022RAA....22l4004N}, FRB 20220912A \citep{2023ApJ...955..142Z}, and FRB 20240114A \citep{2025arXiv250714707Z, 2026ApJS..284...77W, 2026SCPMA..6949512Z, 2026ApJ...998..276Z}, observed by FAST under the FAST FRB Key Science Project. All the observations are carried out with the central beam of the L-band 19-beam receiver of FAST \citep{2019SCPMA..6259502J}. The FRB 20201124A sample consists of observations from its first two active episodes, which are combined and treated as a single dataset in our analysis.
These data are first searched for single pulse candidates, and then the candidates are manually identified before further processing. For the confirmed bursts, the dispersion measure is further refined, followed by polarization calibration to correct for instrumental effects using the noise diode at the beginning of each observation.
In most of the radio emissions, the dominant propagation effect is the Faraday rotation, which generates a $\lambda^2$ dependent phase due to the velocity difference of right and left circularly polarized waves. In a strong magnetic field, the $\lambda^2$ dependence can be broken, and Faraday conversion would appear \citep{2025ApJ...988..164W}.
The Faraday rotation effect is orthogonal to cosmic birefringence because Faraday rotation scales as $\lambda^2$, whereas cosmic birefringence is expected to be frequency independent.
Therefore, the observed PA would be a function of frequency under the consideration of the Faraday rotation effect only,
\begin{equation}
    \mathrm{PA}_{\mathrm{obs}} = \mathrm{PA}_0 + \mathrm{RM} \cdot \lambda^2,
\end{equation}
where $\mathrm{PA}_0$ is the intrinsic PA of astronomical light, and RM is the rotation measure.
The RM synthesis \citep{2005A&A...441.1217B} or the QU-fitting method \citep{2019Sci...365.1013D} is employed on the calibrated Stokes data to determine the RM of each burst. Then the Stokes $Q$ and $U$ are derotated by this RM and averaged over all frequency channels to get the derotated $Q$ and $U$ profiles. The intrinsic PA can be expressed as
\begin{equation}
    \mathrm{PA} = \frac{1}{2} \arctan \left( \frac{U_{\mathrm{derot}}}{Q_{\mathrm{derot}}} \right),
\end{equation}
and we propagate the noise root mean square of the neighboring region to get the uncertainty in PA
\begin{equation}
    \sigma_{\mathrm{PA}} = \frac{1}{2} \frac{Q_{\mathrm{derot}} U_{\mathrm{derot}}}{Q_{\mathrm{derot}}^2 + U_{\mathrm{derot}}^2} \sqrt{\left(\frac{\sigma_{Q_{ \mathrm{derot}}}}{Q_{\mathrm{derot}}}\right)^2 + \left(\frac{\sigma_{U_{\mathrm{derot}}}}{U_{\mathrm{derot}}}\right)^2}.
\end{equation}
Since polarimetric analysis usually requires a high signal-to-noise ratio (SNR) threshold to avoid the illusion, only bursts with SNR$>$50 are retained in our sample, and PA points with $\sigma_{\mathrm{PA}} > 5$ deg are discarded.
For a more detailed description of these processing pipelines and population properties of bursts, you can find them in the corresponding literature \citep{2022Natur.609..685X, 2022RAA....22l4003J, 2023ApJ...955..142Z, 2026arXiv260320663W}.

The collected PA samples exhibit a diverse range of PA swings \citep{2025ApJ...988..175L}. While the majority of bursts show nearly constant PA across the burst profile, a minority display smooth PA swings or irregular variations. In this work, we characterize each burst using a single mean PA, $\overline{\mathrm{PA}}$, this approximation that is well suited for bursts with approximately constant PA but may be inadequate for those exhibiting significant PA evolution. To assess the robustness of our results against this simplification, we construct a golden sample consisting of bursts whose PA curves are well described by a constant value.
Specifically, we select bursts with a reduced chi-square of a constant-PA fit, $\chi^2_{\nu} = \chi^2_{\mathrm{min}}/(N-1) < 5$, or a maximum PA excursion smaller than 20 deg. This subsample provides a cleaner dataset with minimal impact from intrinsic model mismatch.

\section{Simulation-based Dark Matter density Estimation}
\label{sec:dm}
In this section, we briefly review the FRB host properties, introduce the simulation-based scaling relations for halo properties, and estimate the DM density using a simulation-based gNFW profile.
\subsection{FRB host properties}
For the ALDM constraint, both the host-galaxy properties and the precise localization of FRB sources are crucial for estimating the dark matter density at the source location.
FRB 20201124A is hosted by a massive star-forming galaxy with a stellar mass of $3 \times 10^{10} ~M_{\odot}$ and a star formation rate (SFR) of $3.4 ~(0.3) ~M_{\odot}\mathrm{yr}^{-1}$ \citep{2022MNRAS.513..982R, 2021ApJ...919L..23F, 2022Natur.609..685X}. The value reported in the parentheses is the corresponding uncertainty. The host galaxy has a redshift of 0.09795, corresponding to a luminosity distance of 453.3 Mpc under the standard Planck cosmology \citep{2020A&A...641A...6P}. Subsequent observations with the European Very Long Baseline Interferometry Network (EVN) localized the source to milliarcsecond precision, yielding a projected offset of 1.3 kpc from the host-galaxy center \citep{2022ApJ...927L...3N}.
FRB 20220912A was first localized by the Deep Synoptic Array (DSA-110) to a host galaxy at a redshift of 0.0771 (362.4 Mpc). The host has a stellar mass of $\log_{10}(M_*/M_{\odot}) = 10.0 ~(0.1)$ and an SFR of $> 0.1 ~M_{\odot} \mathrm{yr}^{-1}$ \citep{2023ApJ...949L...3R}. Follow-up EVN observations measured a projected offset of 0.8 kpc from the host-galaxy center \citep{2024MNRAS.529.1814H}.
FRB 20240114A is associated with a dwarf galaxy at a redshift of 0.1306 (633.9 Mpc).
The host has a relatively low stellar mass of $\log_{10}(M_*) = 8.6~(0.1)$ and an SFR of 0.06~(0.01) $M_{\odot}\mathrm{yr}^{-1}$. The source is located only 0.5 kpc from the galaxy center \citep{2025ApJ...992L..35B, 2025ApJ...980L..24C}. Besides this, the central galaxy orbited by the host has a larger stellar mass of $\log_{10}(M_*/M_{\odot}) = 9.8~(0.1)$ and an SFR of $\sim 0.16 ~M_{\odot} \mathrm{yr}^{-1}$. The offset of the host from the center of the central galaxy is quite large, reaching up to 82 kpc. This leads to a negligible contribution to the DM density budget of FRB 20240114A.

Actually, the reported offset corresponds to the projected (transverse) separation from the host galaxy center and therefore provides a lower limit on the true three-dimensional distance. Using the projected offset may consequently lead to a slight overestimate of the local dark-matter density. However, FRBs are generally expected to be associated with the luminous part of their host galaxies rather than being randomly distributed throughout the extended DM halo. Since the stellar and star-forming material is strongly concentrated toward the central regions of galaxies, large line-of-sight displacements are statistically disfavored. Therefore, the true spatial offset is expected to be comparable to the observed projected offset, implying that projection effects introduce only a modest systematic uncertainty in the inferred DM density.

\subsection{Estimation from the simulation}
The precise estimation of the local DM density relies on the measurement of the Milky Way rotation curve. At the same time, the host galaxies of the FRB sources usually lack related observations. This motivates us to find an alternative way to determine it, for example, through a self-consistent cosmological simulation. \texttt{UNIVERSEMACHINE} \citep{2019MNRAS.488.3143B} is an empirical galaxy-halo connection model that statistically links galaxy star-formation histories to the assembly histories of their host dark matter halos. By optimizing the model with a wide range of observations across cosmic time, it provides realistic predictions for galaxy properties and their evolution within cosmological simulations. 

The Data Release 1 of \texttt{UNIVERSEMACHINE} provides galaxy catalogs at a dense grid of redshift snapshots. For each FRB host galaxy, we first select the snapshot closest to its observed redshift. We then perform a conditional matching procedure by selecting simulated galaxies whose stellar masses and SFR are consistent with the observed host properties at a 1-$\sigma$ confidence level. The virial halo mass and virial radius of the matched galaxies are adopted as posterior samples of the host halo properties, thereby naturally accounting for the intrinsic scatter in the stellar-halo mass relation.

The inner slope of the gNFW profile can be obtained from the fitting formula of the star formation efficiency $x=M_{*}/M_{\mathrm{vir}}$,
\begin{equation}
    \gamma (X) = n - \log_{10} \left[ n_1\left(1+\frac{X}{x_1}\right)^{-\beta} + \left(\frac{X}{x_0}\right)^{\zeta} \right],
\end{equation}
where the fitting parameters $n = -0.158$, $n_1 = 26.49$, $x_0 = 8.77 \times 10^{-3}$, $x_1 = 9.44 \times 10^{-5}$, $\beta = 0.85$, and $\zeta = 1.66$ are adopted from \cite{2016MNRAS.456.3542T}.
The concentration mass relation can be modeled as a power law form,
\begin{equation}
    \log _{10} c=a+b \log _{10}\left(M_{\mathrm{vir}} /\left[10^{12} h^{-1} M_{\odot}\right]\right),
\end{equation}
where $a = 1.025$, $b = -0.097$, and $h=0.67$ is the dimensionless Hubble parameter \citep{2014MNRAS.441.3359D}. We assume a log-normal scatter of 0.11 dex in halo concentration at fixed halo mass. With these parameters determined, the DM density at the FRB offset can be directly evaluated using the gNFW profile.

\subsection{The gNFW halo profile}
We model the DM halo of the host galaxies using the gNFW profile \citep{1996MNRAS.278..488Z}, which extends the standard NFW profile \citep{1997ApJ...490..493N} by allowing the inner density slope $\gamma$ to vary. The DM density profile is given by
\begin{equation}
    \rho (r) = \frac{\rho_c}{(r/r_s)^\gamma(1+r/r_s)^{3-\gamma}}
\end{equation}
where $\gamma$ is the inner slope, $\rho_c$ is the characteristic density, and $r_s = r_{\mathrm{vir}}/c$ is the scale radius, where $r_{\mathrm{vir}}$ and $c$ are the virial radius and concentration parameter, respectively.
The density follows $\rho(r) \propto r^{-\gamma}$ in the inner region $r \ll r_s$, and $\rho(r) \propto r^{-3}$ in the outer region $r \gg r_s$.
The determination of the virial halo mass can be obtained by integrating this profile to get the virial halo mass $M_{\mathrm{vir}}$, and $\rho_c$ can be expressed as
\begin{equation}
    \rho_c = \frac{M_{\mathrm{vir}}}{4 \pi r_s^3 \int_0^{c} x^{(2-\gamma)} (1+x)^{(\gamma-3)} \mathrm{d}x}.
\end{equation}
Therefore, the unique gNFW profile can be determined from a set of parameters \{$M_{\mathrm{vir}}$, $r_{\mathrm{vir}}$, $c$, $\gamma$\}.

In summary, we generate realizations of host-galaxy and halo properties based on simulated galaxy catalogs and empirical scaling relations, conservatively estimate the corresponding DM density profiles, and derive the source DM density realization from its offset within the host halo.

\subsection{DM density of FRB sources}
\begin{figure}[htbp]
\centering
\includegraphics[width=.8\textwidth]{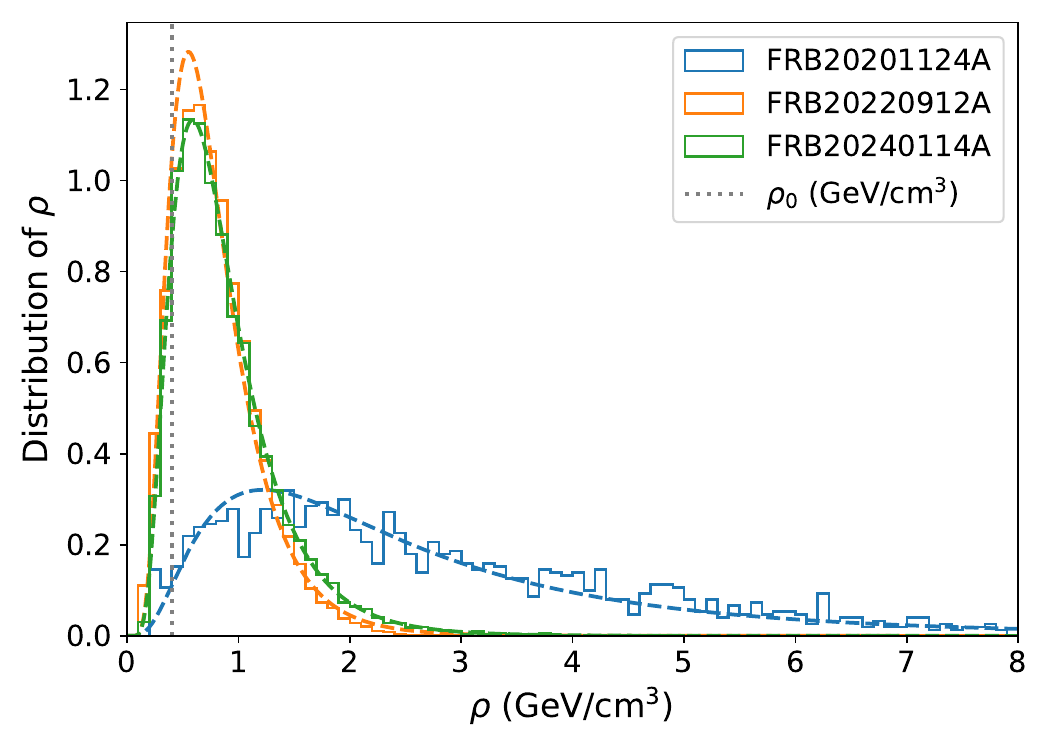}
\caption{The DM density distributions of FRB 20201124A, FRB 20220912A, and FRB 20240114A estimated from our the simulation-calibrated gNFW profile. The histograms show the DM density distributions, and the dashed line with the same color is the corresponding best-fit lognormal curve. The dotted grey line shows the local DM density $\rho_0=0.4~\mathrm{GeV/cm^3}$ \citep{2010A&A...523A..83S}.\label{fig:DMdensity_fit}}
\end{figure}

The DM density distributions of different FRB sources and the corresponding lognormal fits are shown in Figure \ref{fig:DMdensity_fit}. For FRB 20201124A, the simulated DM density distribution is well described by a lognormal distribution with $\rho \sim \mathcal{LN}(0.78, 0.76)$. The corresponding median density is 2.18 $\mathrm{GeV/cm^3}$, which is higher than the local DM density 0.4 $\mathrm{GeV/cm^3}$ with a wide distribution. For FRB 20220912A, we obtain the fitting result $\rho \sim \mathcal{LN}(-0.35, 0.50)$ and the corresponding median density is 0.70 $\mathrm{GeV/cm^3}$.
Despite residing in the lowest stellar mass host galaxy, FRB 20240114A exhibits the high median DM density $\sim 0.77$ $\mathrm{GeV/cm^3}$, and the fit of the lognormal distribution yields $\rho \sim \mathcal{LN}(-0.26, 0.52)$. In addition, we also verify our method using the stellar mass $\log_{10}(M_{*}/M_{\odot})=10.7$ and $\mathrm{SFR} = 1.65 ~M_{\odot} \mathrm{yr}^{-1}$ of the Milky Way \citep{2015ApJ...806...96L}, and the offset 8.2 kpc of the Solar System \citep{2019A&A...625L..10G} to estimate the local DM density, and the inferred result is in good agreement with the current value 0.3-0.4 $\mathrm{GeV/cm^3}$ \citep{2010A&A...523A..83S}.

The inferred DM density of FRB 20220912A is about 0.7 $\mathrm{GeV/cm^3}$, which is much lower than 6.1 $\mathrm{GeV/cm^3}$ reported in previous work \citep{2025CmPhy...8..130W}. 
This discrepancy mainly arises from the
different construction of the host-galaxy analogs. The previous study adopted the best-fitting DM profile of a single galaxy, NGC~4451, as a
proxy for the FRB host. Although NGC~4451 has a comparable stellar mass, the structural comparison is not fully consistent: its quoted $2.2~\mathrm{kpc}$ scale is the maximum radial extent of the CO rotation curve \citep{2022MNRAS.512.1012C}, whereas the corresponding value for the FRB host is its effective radius. Moreover, the highly concentrated halo inferred for this individual galaxy is not necessarily representative of the broad distribution of halo masses, concentrations, and inner density profiles at fixed stellar mass. Our analysis uses a population of cosmological analogs selected with consistently defined stellar masses and SFRs and marginalizes over the intrinsic halo-to-halo scatter, leading to the lower and statistically more representative local DM density reported here.

From these inferred DM densities of different FRB sources, it is worth noting that all these values are comparable to the local DM density, so the ALDM-induced PA oscillation in Eq.~\eqref{eq:aldm} can not be approximated as a simple form dominated by the source or Earth terms. Therefore, the full expression of the superposition of two oscillatory components should be used to determine the ALDM properties.

\section{Constraints on the Chern-Simons Coupling}
\label{sec:res}
In this section, we present the results of the individual-source and FRBPA analyses and compare the resulting constraints with those from other axion searches. We then discuss the advantages and limitations of using repeating FRBs to probe axion dark matter.

\subsection{The constraining power of repeating FRBs}
\begin{figure}[htbp]
\centering
\includegraphics[width=1.0\textwidth]{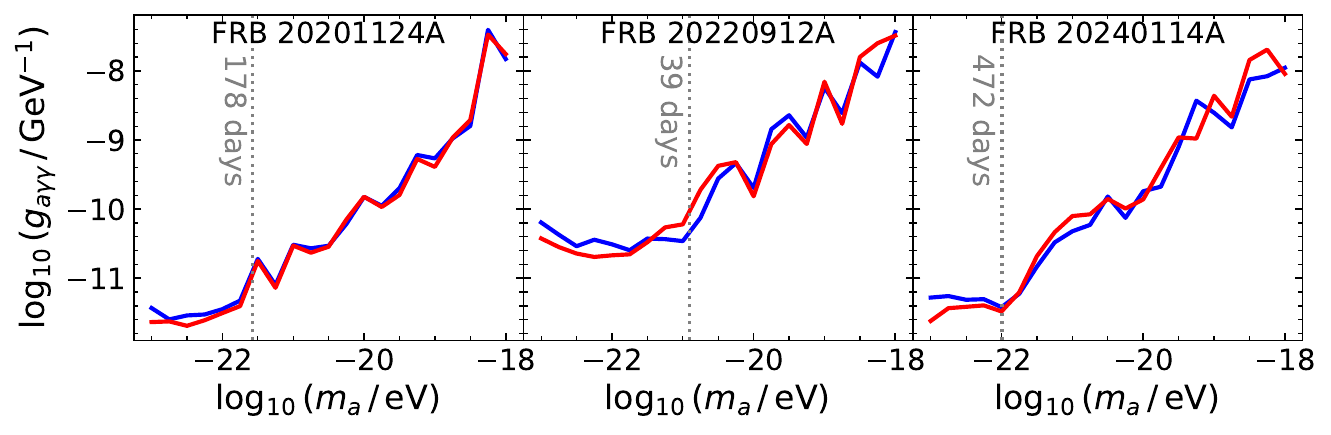}
\caption{95\% upper limit inferred from FRB 20201124A, FRB 20220912A and FRB 20240114A datasets for the axion mass range $[10^{-23},~10^{-18}]$ eV are shown in the panels. The blue and red lines represent the results of the full and golden samples, respectively. The grey dotted line shows the observation period of each dataset.\label{fig:ALPlimit_of_rFRBs}}
\end{figure}

The constraints on the Chern-Simons coupling strength $g_{a\gamma\gamma}$ of three individual repeating FRBs are shown in Figure \ref{fig:ALPlimit_of_rFRBs}. The observation periods of FRB 20201124A, FRB 20220912A, and FRB 20240114A are 178, 39, and 472 days, respectively.
Due to the linear relation between $m_a$ and $g_{a\gamma\gamma}$, the constraints on the lower mass would be stricter. However, when the oscillation period exceeds the observation period, the slowly varying ALDM signal then becomes increasingly degenerate with the time-independent mean PA, making the actual upper limit of $g_{a\gamma\gamma}$ weaker than the linear prediction $g_{a\gamma\gamma} \propto m_a$.
Consequently, the constraint ceases to improve, and the exclusion curve gradually flattens toward lower masses.
For all datasets, the full and golden samples nearly have the same constraining power, and the linear relation is also maintained when the ALDM-induced PA oscillating period is shorter than the observation period.
Compared with the previous results based on the same FRB~20220912A dataset \citep{2025CmPhy...8..130W}, our constraints are broadly comparable over the overlapping mass range but become more stringent at certain axion masses, while also extending the search to a broader mass
range. This improvement is primarily driven by our hierarchical Bayesian framework, which properly accounts for the intrinsic burst-to-burst PA scatter and better distinguishes it from the ALDM-induced birefringence signal.

\subsection{The constraining power of FRBPA}
\begin{figure}[htbp]
\centering
\includegraphics[width=.8\textwidth]{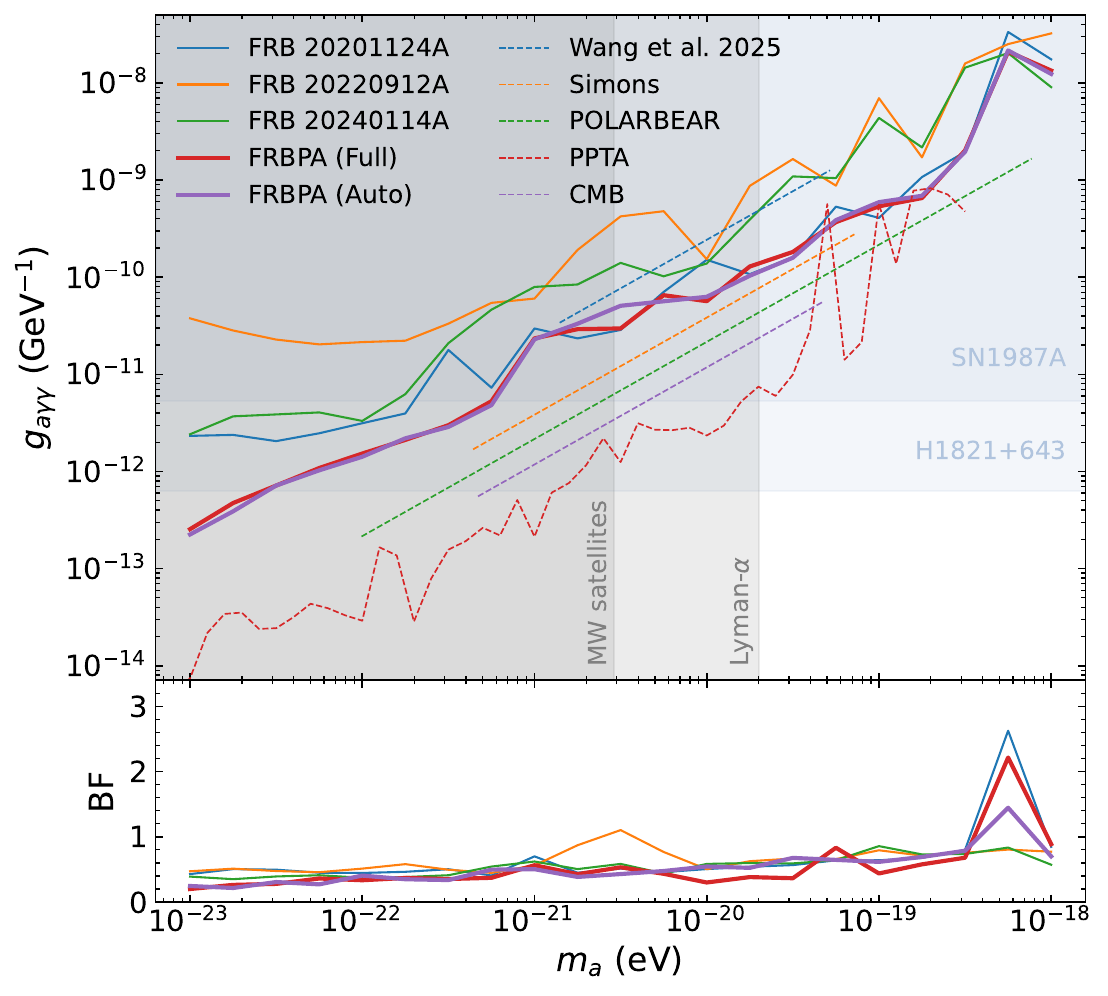}
\caption{95\% upper limits of the repeating FRB PA datasets on the axion-photon coupling $g_{a\gamma\gamma}$ as a function of the axion mass $m_a$ and their corresponding BF are shown in the upper and lower panels, respectively. The blue, orange, green, and red lines represent the constraints from FRB 20201124A1, FRB 20201124A2, FRB 20220912A, and FRB 20240114A, respectively. The smoothed constraint of the same dataset of FRB 20220912A from \cite{2025CmPhy...8..130W} is also shown in this figure. As a reference, we show the smoothed constraints from the observation of PA variation in Crab Nebula with POLARBEAR \citep{2024PhRvD.110f3013A} and Simons Array \citep{2026PhRvD.113d3044A}. The grey dashed line is the limit reported by CMB polarization \citep{2022PhRvD.106d2011F}. The pink dashed line represents the constraints from the spatial correlation of the Parkes Pulsar Polarization Array \citep{2026PhRvL.136a1001X}. We also show the upper limits from gamma-rays of SN 1987A \citep{2015JCAP...02..006P}, and X-ray spectral distortions of quasar H1821+643 \citep{2022MNRAS.510.1264S}. The grey regions show the constraints of the axion mass from Milky Way satellites \citep{2021PhRvL.126i1101N} and Lyman-$\alpha$ forest \citep{2021PhRvL.126g1302R}.\label{fig:comparison}}
\end{figure}

In the construction of FRBPA, we only use the golden sample of these three repeating FRB sources.
The upper panel of Figure \ref{fig:comparison} compares the constraints obtained from the individual sources and combined FRBPA analyses with existing limits from other astrophysical probes.
Compared to the results from individual sources, the most notable improvement provided by the FRBPA occurs at the low-mass end. In the ALDM mass range of $10^{-23}\,\mathrm{eV}\lesssim m_a\lesssim10^{-22}\,\mathrm{eV}$, the full-correlation FRBPA strengthens the upper limit on $g_{a\gamma\gamma}$ by approximately a factor of 2-10 relative to the strongest constraint from individual sources.
This demonstrates the advantage of jointly combining the PA measurements and complementary observing spans of multiple repeating FRBs. Toward higher axion masses, the improvement becomes less pronounced, and the FRBPA constraints are generally comparable to those obtained from the individual sources.
The full-correlation and the auto-correlation FRBPA analyses yield similar constraints over the considered ALDM mass region, which suggests that the current sensitivity of FRBPA is primarily driven by the auto-correlation, while the common Earth term provides a limited contribution.
The FRBPA limits are comparable to the upper limit from Simons Observatory \citep{2026PhRvD.113d3044A}, although they remain weaker than several of the other astrophysical limits shown in the figure.

The bottom panel of Figure \ref{fig:ALPlimit_of_rFRBs} shows the BFs comparing the ALDM and null hypotheses for the individual-source and FRBPA analyses. The BFs remain below unity over most of the mass range, indicating that the data generally favor the null hypothesis. The BF curves for FRB 20201124A and the FRBPA exhibit modest peaks near $m_a=5.6\times10^{-19}\,\mathrm{eV}$ and the corresponding ALDM oscillating period is about $2.05$ h, which is very close to the observation length 2 h during the first active episode of FRB 20201124A.
This peak appears only in the datasets associated with FRB 20201124A, suggesting that it is likely an artifact associated with the finite observation length rather than a physical feature.
Moreover, the maximum value of BF remain below $10^{0.5}$, corresponding to negligible evidence for the ALDM hypothesis on the Jeffreys scale. We therefore find no statistically significant evidence for an ALDM-induced periodic PA modulation in any of the three repeating FRBs or in the combined FRBPA analysis. The resulting constraints are consequently interpreted as upper limits on the axion-photon coupling.

\begin{figure}[htbp]
\centering
\includegraphics[width=1.0\textwidth]{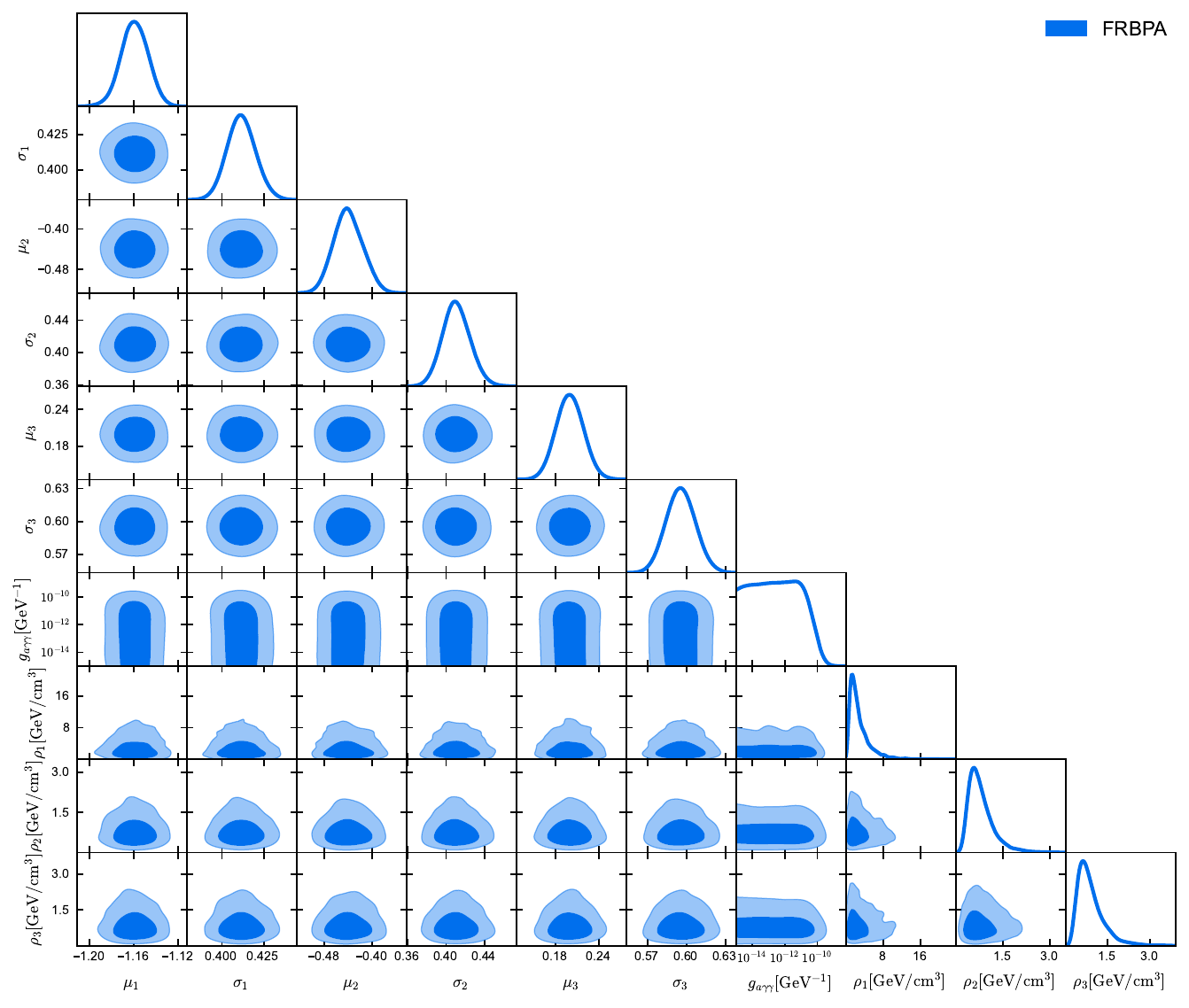}
\caption{1D and 2D marginalized posterior distributions from the full-correlation FRBPA analysis for an ALDM mass of $m_a = 10^{-20}$ eV. The shaded regions in the 2-D panels denote the 68.3\% and 95.4\% credible regions. The subscripts 1, 2, and 3 correspond to FRB 20201124A, FRB 20220912A, and FRB 20240114A, respectively. \label{fig:MCMCexample}}
\end{figure}

\subsection{Discussion}
There is a key difference between FRB-based probes and other ALDM probes. Unlike most existing probes, FRBs are sufficiently bright that the statistical uncertainty associated with individual PA curves is typically negligible compared to the intrinsic scatter of the PA distribution. As a result, the sensitivity is governed primarily by three factors: the intrinsic variance of the PA distribution, the number of detected bursts, and the coverage of the axion-induced phase evolution. A smaller intrinsic variance provides a cleaner baseline against which a periodic modulation can be identified, while a larger burst sample reduces statistical fluctuations and improves the reconstruction of the underlying signal.
An important limitation arises from the highly variable activity of
repeating FRBs.
Bursts are often clustered within short active episodes,
separated by long periods of relative quiescence. This irregular sampling may leave substantial portions of an ALDM oscillation cycle uncovered, particularly when the oscillation period is comparable to or longer than the observing span. The resulting incomplete phase coverage makes the slowly varying ALDM signal more difficult to distinguish from the time-independent mean PA and therefore reduces the sensitivity in the low-mass regime.

Our results demonstrate that combining multiple repeating FRBs into the FRBPA can substantially improve the constraints at the low-mass end. By jointly analyzing sources with different activity windows and observing spans, the FRBPA provides more complete temporal sampling and reduces the dependence of the constraints on the activity pattern of any individual source. In addition, the full-correlation analysis can exploit the cross-correlations induced by the common Earth term. These advantages motivate the development of larger FRBPAs as more repeating FRBs become available.

Future interferometric transient surveys with daily monitoring cadence, such as the Canadian Hydrogen Observatory and Radio-transient Detector (CHORD; \cite{2019clrp.2020...28V}) and the Square Kilometre Array (SKA; \cite{2009IEEEP..97.1482D, 2023SCPMA..6620412Z, 2020MNRAS.497.4107H, 2026arXiv260627714C}), will enable continuous monitoring of repeating FRBs over many years, providing both dense phase sampling and long observational baselines. These two factors are crucial for improving sensitivity to ultra-light axions, whose oscillation periods can be comparable to or longer than current monitoring spans. Moreover, such surveys are expected to substantially increase the burst samples of known repeaters and discover many new active sources. Persistently active repeaters with stable polarization position angles will therefore constitute ideal laboratories for testing ALDM-induced cosmic birefringence in the coming decade.

\section{Conclusion}
\label{sec:con}
In this work, we develop a hierarchical Bayesian framework to search for ALDM-induced cosmic birefringence at the population level using PA measurements of repeating FRBs.
By jointly modeling the intrinsic burst-to-burst PA scatter and the temporal modulation expected from oscillating ALDM, this framework enables robust inference of the Chern-Simons coupling from FRB datasets. We further extend the framework to an FRBPA, which combines multiple repeating FRB sources and substantially strengthens the constraints, particularly in the low-mass regime.

To accurately model the expected birefringence signal, we construct a framework that connects host galaxy observations to the DM density at the FRB source. By combining the stellar-halo mass relation from the simulated galaxy catalog, galaxy-halo scaling relations, and gNFW DM halo profile, we infer the DM density of the FRB source conservatively. This allows us to estimate the source contribution to the birefringence signal and its associated uncertainty. We find that, for repeating FRBs considered in this paper, the source and Earth contributions are generally of comparable magnitude, implying that both terms must be retained in a self-consistent analysis. Neglecting either contribution can lead to biased predictions of the signal amplitude and consequently of the inferred axion constraints.

We apply our hierarchical Bayesian framework to three large datasets observed by FAST, namely FRB 20201124A, FRB 20220912A, and FRB 20240114A. These sources provide large samples of polarized bursts and observation baselines ranging from several days to more than one year, enabling a search for periodic polarization variations induced by ALDM over a broad mass range. Then these three samples are combined into a single sample to form an FRBPA, which improves the poor phase coverage of individual sources and significantly enhances the sensitivity at the low-quality end.
No statistically significant evidence for an axion-induced birefringence signal is found in either the individual dataset or FRBPA analyses. We therefore derive upper limits on the axion-photon coupling $g_{a\gamma\gamma}$ across the accessible axion mass range.

Our results demonstrate that repeating FRBs provide a promising probe of ultralight ALDM. Unlike many existing searches, the uncertainty of individual polarization measurements is typically negligible compared with the intrinsic burst-to-burst scatter of the PA distribution. As a result, the sensitivity is governed primarily by the intrinsic variance of the PA distribution, the number of detected bursts, and the temporal coverage of the axion-induced phase evolution.
With the rapidly growing sample of repeating FRBs and increasing size of single source sample, the methodology developed in this work can be readily extended to future observations, offering substantially improved sensitivity to ALDM-induced cosmic birefringence.

\acknowledgments
\begin{CJK*}{UTF8}{gbsn}
This work made use of data from the FAST FRB Key Science Project.
X.-H. L. acknowledges support from the National SKA Program of China （Nos. 2022SKA0110100 and 2022SKA0110101) and NSFC (grant No. 12361141814).
Z.-Y. Y. is supported by an appointment to the Young Scientist Training (YST) program at the APCTP through the Science and Technology Promotion Fund and Lottery Fund of the Korean Government. 
W.-Y. W. acknowledges support from the NSFC (No.12261141690 and No.12403058), the National SKA Program of China (No. 2020SKA0120100), and the Strategic Priority Research Program of the CAS (No. XDB0550300).
X.-L. C. acknowledges support from the NSFC (grant No. 12361141814, 12421003), and by the Specialized Research Fund for State Key Laboratory of Radio Astronomy and Technology.
\end{CJK*}
% \paragraph{Note added.} This is also a good position for notes added
% after the paper has been written.

% Bibliography

%% [A] Recommended: using JHEP.bst file
%% \bibliographystyle{JHEP}
%% \bibliography{biblio.bib}

%% or
%% [B] Manual formatting (see below)
%% (i) We suggest to always provide author, title and journal data or doi:
%% in short all the informations that clearly identify a document.
%% (ii) please avoid comments such as "For a review'', "For some examples",
%% "and references therein" or move them in the text. In general, please leave only references in the bibliography and move all
%% accessory text in footnotes.
%% (iii) Also, please have only one work for each \bibitem.

\bibliographystyle{JHEP}
\bibliography{biblio}

\end{document}